\documentclass[epj,final]{svjour}
\usepackage{graphicx,times}

\begin{document}

%
%==============================================================================
% Title
%==============================================================================
%
\title{\boldmath%
 Strange nucleon form factors: Solitonic approach to\\
 $G_{M}^{s}$, $G_{E}^{s}$, $\tilde{G}_{A}^{p}$ and $\tilde{G}_{A}^{n}$
 and comparison with world data.\unboldmath}

\titlerunning{Strange nucleon form factors}

\author{%
 Klaus Goeke\inst{1}\and
 Hyun-Chul Kim\inst{2}\and
 Antonio Silva\inst{3, 4}\and
 Diana Urbano\inst{3, 4}}

\authorrunning{K.~Goeke {\it et al.}}

\institute{%
 Institut f{\"u}r Theoretische Physik II, Ruhr-Universit{\"a}t Bochum,
 D--44780 Bochum, Germany\and
 Department of Physics and Nuclear Physics \& Radiation Technology
 Institute (NuRI), Pusan National University,\\
 609-735 Busan, Republic of Korea\and
 Departamento de F{\'i}sica and Centro de F{\'i}sica Computacional,
 Universidade de Coimbra, P-3000 Coimbra, Portugal\and
 Faculdade de Engenharia da Universidade do Porto, R. Dr. Roberto Frias
 s/n, P-4200-465 Porto, Portugal}

\date{August 2006}

\abstract{%
 We summarize the results of the chiral quark-soliton model
 ($\chi$QSM) concerning basically all form factors necessary to
 interpret the present data of the parity violating electron
 scattering experiments SAMPLE, HAPPEX, A4 and G0. The results
 particularly focus on the recently measured asymmetries and the
 detailed data for various combinations of $G_{M}^{s}$,
 $G_{E}^{s}$, $\tilde{G}_{A}^{p}$ and $\tilde{G}_{A}^{n}$ at
 $Q^2=0.1$ GeV$^2$. The calculations yield positive strange magnetic
 and electric form factors and a negative axial vector one, all being
 rather small. The results are very close to the combined
 experimental world data from parity violating electron scattering
 and elastic $\nu p$- and $\bar{\nu p }$- scattering.}


\maketitle

%
%==============================================================================
% Section 1
%==============================================================================
%
\textbf{1.} The strange quark contribution to the distributions of
charge and magnetization in the nucleon has been a very important
issue well over decades, since it provides a vital clue in
understanding the structure of the nucleon and in particular in
probing the $\mathrm{q\bar{q}}$-sea. There are some indications of
about 4\% contribution to the momentum sum rule of deep inelastic
lepton scattering, of roughly 15\% to the spin of the nucleon
extracted from polarized deep inelastic scattering, or of up to
30\% contribution to the mass of the nucleon, where all these
numbers show rather large uncertainties. Recently, the strangeness
content of the nucleon has been studied particularly intensively
since parity-violating electron scattering (PVES)
has demonstrated to provide an essential tool for probing the sea of
$\mathrm{s\bar{s}}$ pairs in the vector channel~\cite{Kaplan:1988ku}.
In fact, various PVES experiments have been already conducted from
which the strange vector form factors can be
extracted~\cite{Mueller:1997mt,Spayde:2000qg,SAMPLE00s,Aniol:2000at,%
Maas:2004dh,Maas:2004ta,Maas:2003xp,happex,Aniol:2005zf,Aniol:2005zg,%
Armstrong:2005hs}. The results from the SAMPLE, HAPPEX, PVA4, and
G0 collaborations have shown evidence for a non-vanishing strange
quark contribution to the structure of the nucleon. In particular,
evidence was found that the strange magnetic moment of the proton
is positive~\cite{Aniol:2005zg}, suggesting that the strange
quarks reduce the proton's magnetic moment. This is an unexpected
and surprising finding, since a majority of theoretical studies
favors a negative value. One of the models, which yield a positive
strange magnetic moment of the proton, is the chiral quark soliton
model ($\chi$QSM). It will be used in the present paper to
investigate the form factors $G_{M}^{s}$, $G_{E}^{s}$,
$\tilde{G}_{A}^{p}$ and $\tilde{G}_{A}^{n}$ and to compare them
with world data.

Using the $\chi$QSM the present authors have recently investigated
the set of six electromagnetic form factors ($G_{E,M}^{u,d,s}$)
and three axial-vector ones
($G_{A}^{u,d,s}$)~\cite{Silva:2001st,Silva:2002ej,Silva,Silvaetal}.
The results show a good agreement with the data of the SAMPLE,
HAPPEX, A4 and G0 experiments. This includes parity violating
asymmetries (PVA) which have been measured by the G0 experiment
over a range of momentum transfers in the forward
direction~\cite{Armstrong:2005hs}. We even predicted the PVAs of
the future G0 experiment at backward angles~\cite{Silva:2005qm}.
In the present contribution we perform more detailed comparison
including the most recent data of the HAPPEX experiment on He-4
and the results of the PVAs combined with elastic $\nu p$- and
$\bar{\nu} p$-scattering.

%
%==============================================================================
% Section 2
%==============================================================================
%
\textbf{2.} The $\chi$QSM has been used several times to calculate
strange properties of the nucleon and of hyperons. It is an
effective relativistic quark theory based on the instanton-degrees
of freedom of the QCD vacuum and has been derived from QCD in the
large-$N_c$ limit. In the end it turns out to be the simplest
possible quark theory which allows for spontaneously broken chiral
symmetry. It results in an effective chiral action for valence and
sea quarks both moving in a static self-consistent Goldstone
background
field~\cite{Christov:1996vm,Alkofer:1994ph,Wakamatsu:1990ud}. For
this model it is absolutely natural to have strange quark
contributions to the nucleon. The $\chi$QSM has very successfully
been applied to mass splittings of hyperons, to electromagnetic
and axial-vector form factors~\cite{Christov:1996vm} of the baryon
octet and decuplet and to forward and generalized parton
distributions of the nucleon. With one set of four parameters,
unchanged for years, it reproduces all appropriate observables of
light baryons with an accuracy of (10--30)\%. This parameter set
consists of an effective current mass for up- and down quarks, a
cut-off parameter in the relativistic proper-time regularization
scheme, and a quark-pion coupling constant corresponding to a
constituent mass. These parameters are fitted to the pion decay
constant, the pion mass, and baryonic properties as proton charge
radius and delta-nucleon mass splitting. In addition we assume an
effective current strange quark mass of $180$ MeV. A numerical
iteration procedure yields then the self-consistent mean field
whose lowest states get occupied until baryon number $B=1$ is
reached. The resulting solitonic state is semiclassically rotated
in space and iso-space in order to project on proper spin and
hypercharge quantum numbers.

As far as strange form factors are concerned the formalism used in
the present investigation can be found in the paper of Silva {\it
et al.}~\cite{Silva:2005qm} and references therein. We just
mention for clarity some relations between form factors, which are
often differently denoted in the literature. We refer to the
papers of Musolf {\it et al.}~\cite{Musolf:1993tb}, of Alberico
{\it et al.}~\cite{Alberico:2001sd} and of Maas {\it et
al.}~\cite{Maas:2004ta}. Altogether we have:
\begin{equation}
 \tilde{G}_{A}^{p}=-(1+R_{A}^{1})G_{A}^{(3)}(Q^{2})+R_{A}^{0}+G_{A}^{s}
\end{equation}
\begin{equation}
 \tilde{G}_{A}^{n}=-3(1+R_{A}^{1})G_{A}^{(3)}(Q^{2})+R_{A}^{0}+G_{A}^{s}
\end{equation}
\begin{equation}
 \tilde{G}_{A}^{p}= G_{A}^{e}=-\frac{1}{2}G_{A}^{NC}.
\end{equation}
with the values for the electro-weak radiative corrections~\cite%
{Musolf:1993tb}:
\begin{equation}
R_{A}^{1}=-0.41\pm 0.24,\;\;R_{A}^{0}=0.06\pm 0.14.
\end{equation}

\begin{figure}
\centering
\includegraphics[width=.4\textwidth]{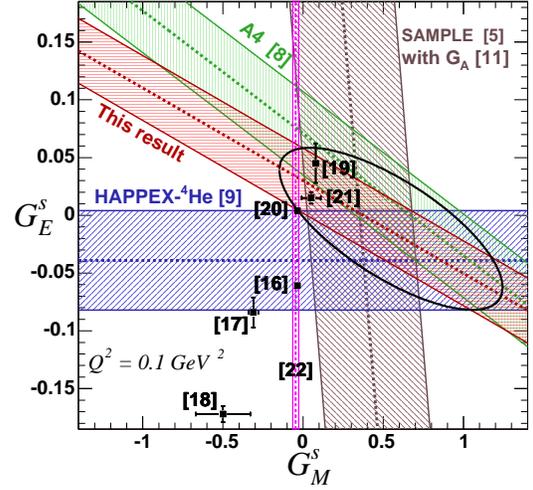}
\caption{The world data on the strange form factors $G_{M}^{s}$
 and $G_{E}^{s}$ at $Q^{2} = 0.1$ GeV$^2$ including the HAPPEX data
 on He-4 of 2004. The figure is taken from
 ref.~\cite{Aniol:2005zg}. The numbers indicate the references of
 theoretical calculations. The $\chi$QSM is given by~\cite{Silva:2001st}.}
\label{fig:milos-GEs-vs-GMs-HAPPEX-early}
\end{figure}

\begin{figure}
\centering
\includegraphics[width=.4\textwidth]{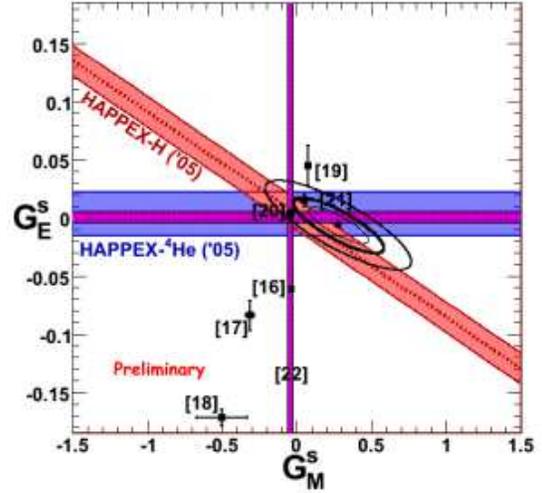}
\caption{The world data on the strange form factors $G_{M}^{s}$
 and $G_{E}^{s}$ at $Q^{2} = 0.1$ GeV$^2$ including the HAPPEX data
 on He-4 of 2004 and of 2005 (preliminary). The numbers indicate
 the references of theoretical calculations. The figure is taken
 from ref.~\cite{Paschke}. The $\chi$QSM is given by~\cite{Silva:2001st}.}
\label{fig:milos-GEs-vs-GMs-HAPPEX-new}
\end{figure}

%
%==============================================================================
% Section 3
%==============================================================================
%
\textbf{3.} The experimental situation is by far the best at
$Q^2=0.1$ GeV$^2$, where in addition to the usual linear combinations
of electric and magnetic form factors the measurements of parity
violation on He-4 allowed an extraction of $G_E^s$. The
experimental results of the HAPPEX Collaboration are
$G_E^s=-0.038 \pm 0.042 \pm 0.010$ measured at $Q^2=0.091$
GeV$^2$~\cite{Aniol:2005zf} and, more recently,
$G_E^s=-0.002 \pm 0.017$ at $Q^2=0.1$ GeV$^2$~\cite{Paschke}.
Also the combined data
$G_E^s=-0.006 \pm 0.016$~\cite{Paschke} are consistent with zero.
Experimental evidence from the SAMPLE and HAPPEX collaborations
gives a positive value of the strange magnetic form factor $G_M^s$
at $Q^2=0.1$ GeV$^2$ of $G_M^s=0.37 \pm 0.20 \pm 0.26 \pm
0.07$~\cite{Spayde:2003nr}, $G_M^s=0.55 \pm
0.28$~\cite{Aniol:2005zg} and $G_M^s=0.12 \pm 0.24$~\cite{Paschke}
respectively. The overall comparison of the $\chi$QSM
calculation~\cite{Silva:2001st} with the world data and with other
model calculations~\cite{Park:1991fb,Hammer:1995de,Hammer:1999uf,%
Lyubovitskij:2002ng,Bijker:2006id}, and lattice gauge
calculations~\cite{Lewis:2002ix,Leinweber:2006ug} is given in
Figure~\ref{fig:milos-GEs-vs-GMs-HAPPEX-early}. If one adds the
preliminary 2005 data of the HAPPEX-He-4 experiment the regions of
confidence get smaller and one obtains the
Figure~\ref{fig:milos-GEs-vs-GMs-HAPPEX-new}. Only those
theoretical calculations are selected which are somehow in the
vicinity of the experimental data. Actually they do not give a
consistent picture. This is also true for lattice-QCD calculations
(LQCD). For example those of Lewis {\it et
al.}~\cite{Lewis:2002ix} advocate a positive magnetic strange
moment, whereas the recent
results of Leinweber {\it et al.}~\cite{Leinweber:2004tc,%
Leinweber:2005bz,Leinweber:2006ug}, indicate a negative one.

%
%==============================================================================
% Figures 1-5
%==============================================================================
%

\vspace{0.5cm}

\begin{figure}
\centering
\includegraphics[width=.4\textwidth]{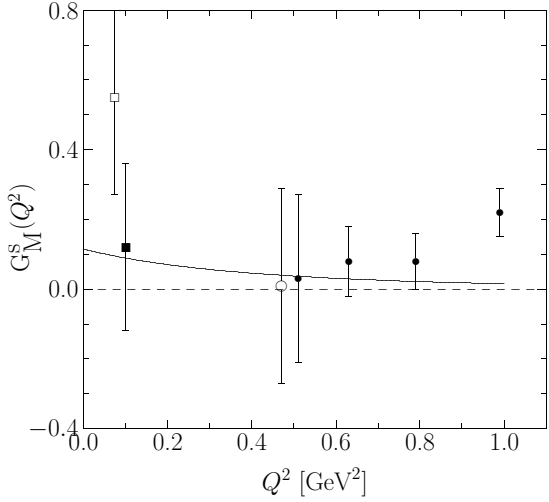}
\caption{The strange magnetic form factor $G_{M}^{s}(Q^2)$ of the
 nucleon: The $\chi$QSM is compared with the analysis of Pate {\it
 et al.}~\cite{Pate:2005bk} involving simultaneously parity
 violating $ep$ data and data from $\nu$- and $\bar{\nu}$-
 scattering. The open circle is from a combination of HAPPEx and
 E734 data, while the closed circles are from a combinaton of G0
 and E734 data. The open square is from ref.~\cite{Aniol:2005zg}
 and involve parity violating $ep$ data only, similarly as the
 closed square from refs.~\cite{Paschke} taken from~\cite{Bijker:2006id}.}
\label{fig:kim-gms-bis-1GeV}
\end{figure}

\begin{figure}
\centering
\includegraphics[width=.4\textwidth]{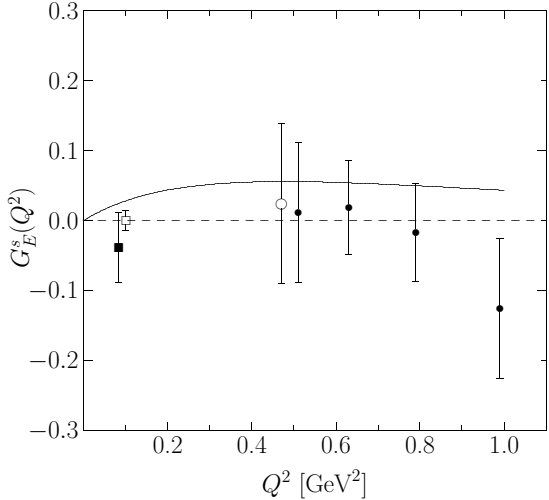}
\caption{The strange electric form factor $G_{E}^{s}(Q^2)$ of the
 nucleon: The $\chi$QSM is compared with the analysis of Pate {\it
 et al.}~\cite{Pate:2005bk}. The open circle is from a combination
 of HAPPEx and E734 data, while the closed circles are from a
 combinaton of G0 and E734 data. The open square is from
 ref.~\cite{Aniol:2005zf} and involve parity violating $ep$ data
 only, similarly as the closed square from refs.~\cite{Paschke}
 taken from~\cite{Bijker:2006id}.}
\label{fig:kim-ges-bis-1GeV}
\end{figure}

\begin{figure}
\centering
\includegraphics[width=.4\textwidth]{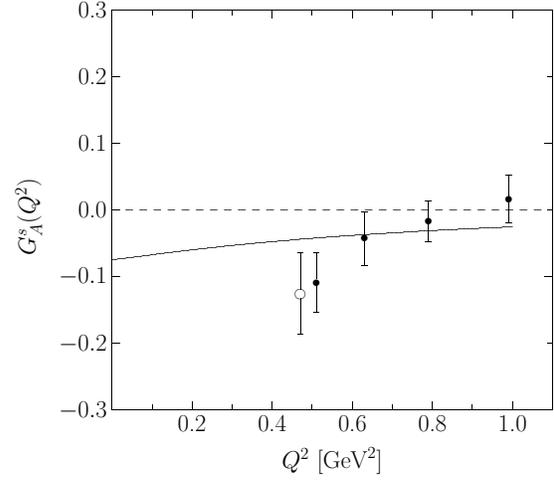}
\caption{The strange axial vector form factor $G_{A}^{s}(Q^2)$ of
the
 nucleon: The $\chi$QSM is compared with the analysis of Pate {\it
 et al.}~\cite{Pate:2005bk}. The open circle is from a combination
 of HAPPEX and E734 data, while the closed circles are from a
 combination of G0 and E734 data.}
\label{fig:kim-gas-bis-1GeV}
\end{figure}

%
%==============================================================================
% Section 4
%==============================================================================
%
\textbf{4.} It is interesting to combine the data from PV electron
scattering with the data from elastic $\nu p$- and $\bar{\nu
p}$-scattering off protons~\cite{Ahrens:1986xe}, which provides
independent information on the strange form
factors~\cite{Pate:2003rk}. A comparison with the results of such
an extraction~\cite{Pate:2005bk} can be seen at
Figures~\ref{fig:kim-gms-bis-1GeV}, \ref{fig:kim-ges-bis-1GeV} and
\ref{fig:kim-gas-bis-1GeV}. Apparently the $\chi$QSM is compatible
with more or less all data available up to $Q^2=1$ GeV$^2$, which
is the range where the $\chi$QSM can provide form factors. It is
not excluded that the experiments favor a negative
$G_{E}^{s}(Q^2)$ whereas the $\chi$QSM yields a positive one.

%
%==============================================================================
% Figures 6-8
%==============================================================================
%
\begin{figure}
\centering
\includegraphics[width=.4\textwidth]{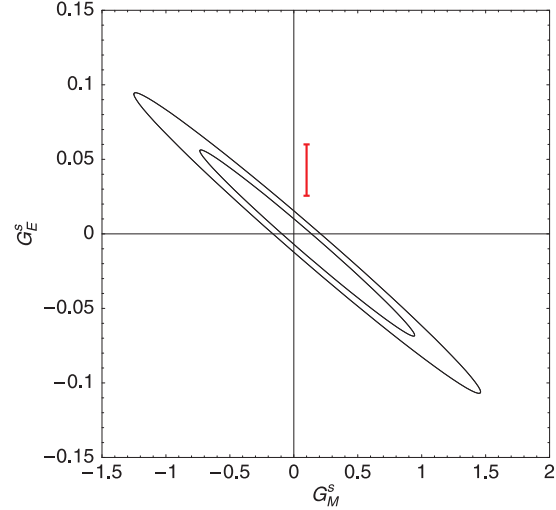}
\caption{The contours display the 68\% and 95\% confidence
 intervals for the joint determination of $G_{M}^{s}$ and
 $G_{E}^{s}$ at $Q^2=0.1$ GeV$^2$. The result of the $\chi$QSM is
 indicated. The ellipses originate from a theory-independent
 combined fit to all parity-violating data by Young
 {\it et al.}~\cite{Young:2006jc}.}
\label{fig:GsE-vs-GsM-young}
\end{figure}

\begin{figure}
\centering
\includegraphics[width=.4\textwidth]{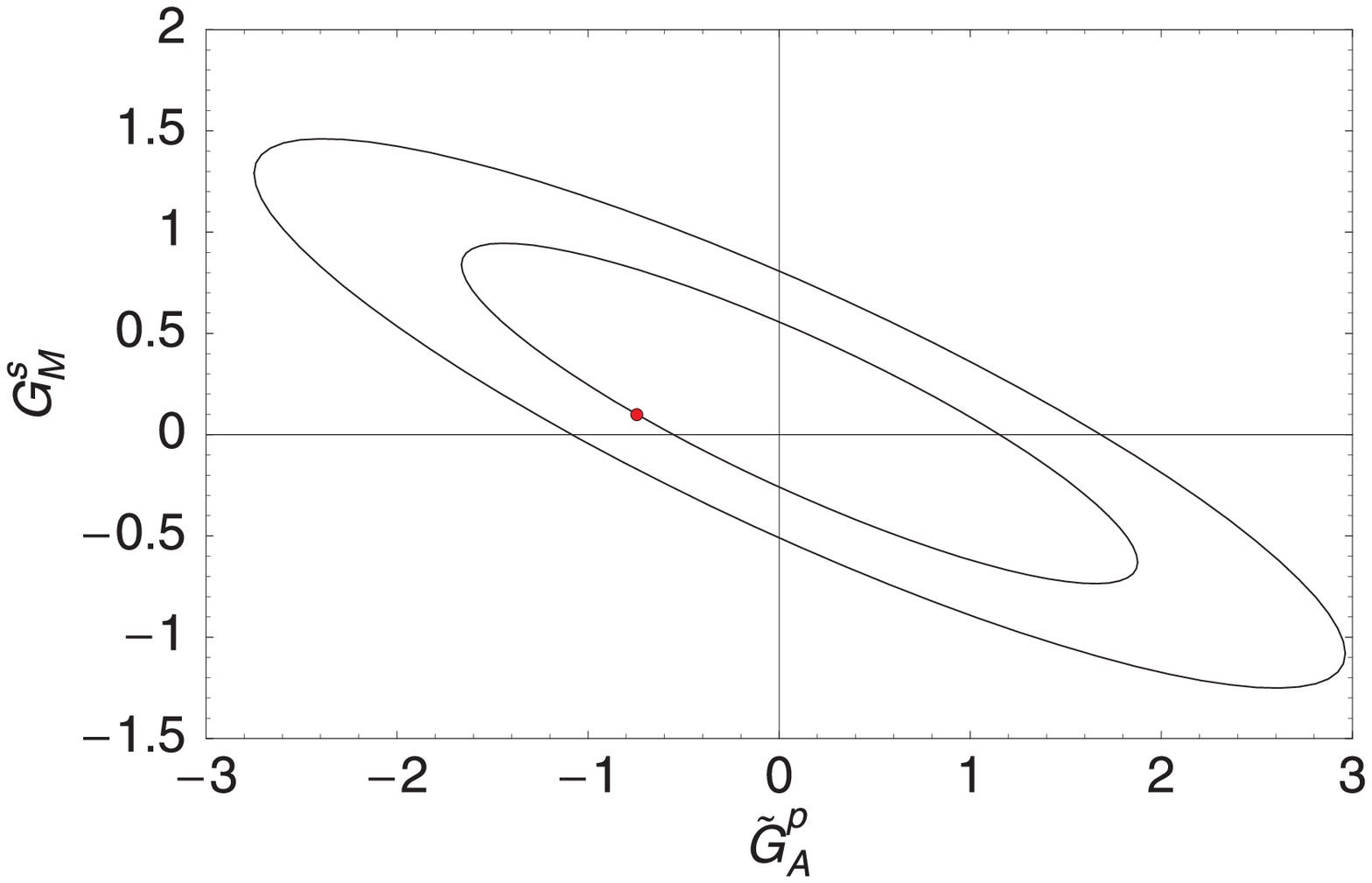}
\caption{The contours display the 68\% and 95\% confidence
 intervals for the joint determination of $G_{M}^{s}$ and
 $\tilde{G}_{A}^{p}$ at $Q^2=0.1$ GeV$^2$. The result of the $\chi$QSM
 is indicated. The ellipses originate from a theory-independent
 combined fit to all parity-violating data by Young
 {\it et al.}~\cite{Young:2006jc}.}
\label{fig:Gsm-vs-Gtildepa}
\end{figure}

\begin{figure}
\centering
\includegraphics[width=.4\textwidth]{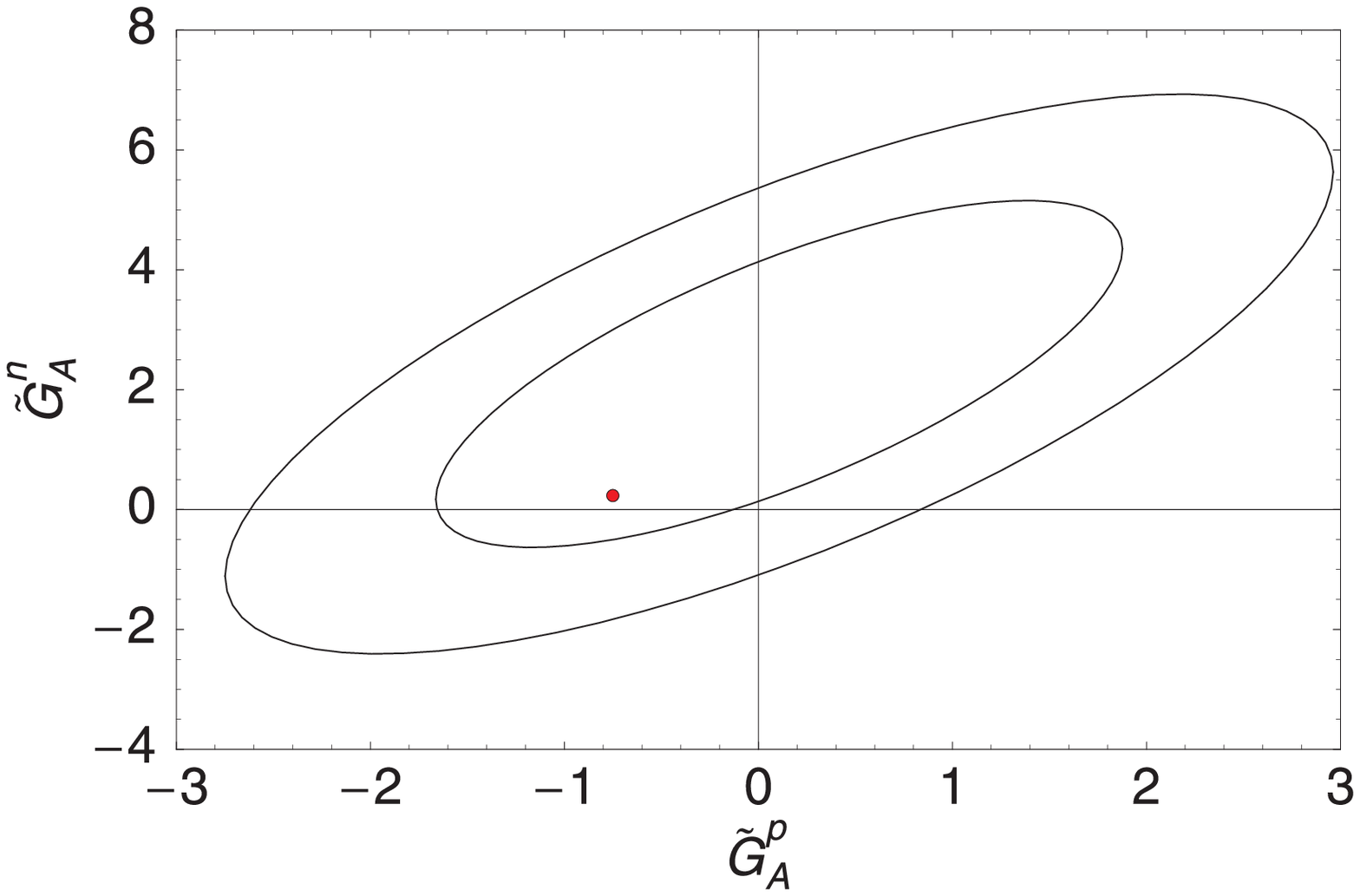}
\caption{The contours display the 68\% and 95\% confidence
 intervals for the joint determination of $\tilde{G}_{A}^{n}$ and
 $\tilde{G}_{A}^{p}$ at $Q^2=0.1$ GeV$^2$. The result of the $\chi$QSM
 is indicated. The ellipses originate from a theory-independent
 combined fit to all parity-violating data by Young
 {\it et al.}~\cite{Young:2006jc}.}
\label{fig:Gntildep-vs-Gntilden}
\end{figure}

%
%==============================================================================
% Section 5
%==============================================================================
%

\textbf{5.} It is also interesting to compare the present
calculations with the analysis of Young {\it et al.} in
ref.~\cite{Young:2006jc}. These authors use systematic expansions
of all the unknown form factors to simultaneously analyze the
current data sets and extract the values at $Q^2=1$ GeV$^2$,
independent of theoretical input, except assuming the constraint
of charge symmetry. Figure~\ref{fig:GsE-vs-GsM-young} shows this
analysis for $G_{M}^{s}$ and $G_{E}^{s}$,
Figure~\ref{fig:Gsm-vs-Gtildepa} for $G_{M}^{s}$ and
$\tilde{G}_{A}^{p}$ and Figure~\ref{fig:Gntildep-vs-Gntilden} for
$\tilde{G}_{A}^{n}$ and $\tilde{G}_{A}^{p}$. The error bar of the
$\chi$QSM-result in Figure~\ref{fig:GsE-vs-GsM-young} is caused by
a systematic error of the model in case of a a purely strange
observable. It originates from the inability of the $\chi$QSM to
describe simultaneously mesonic tails with different Yukawa
masses~\cite{Silva:2001st}. For quantities, which are not purely
strange, this systematic error is usually negligible.

%
%==============================================================================
% Section 6
%==============================================================================
%
\textbf{6.} In the present theoretical work, we have investigated
various form factors which are relevant for the analysis of
parity-violating electron scattering experiments SAMPLE, HAPPEX,
A4 and G0 and the scattering of $\nu$- and $\bar{\nu}$- scattering
off nucleons. These form factors are $G_{M}^{s}$, $G_{E}^{s}$,
$\tilde{G}_{A}^{n}$ and $\tilde{G}_{A}^{p}$. We used for the study
the electromagnetic and strange vector and axial vector form
factors calculated in the chiral quark soliton model ($\chi$QSM),
yielding both small but positive magnetic and electric strange
form factors, see
refs.~\cite{Silva:2001st,Silva:2002ej,Silva,Silvaetal}. All these
$\chi$QSM form factors were obtained with one fixed set of four
model parameters, which has been adjusted several years ago to
basic mesonic and baryonic observables. As seen already in a
previous paper of the present authors~\cite{Silva:2005qm} the
parity-violating asymmetries obtained in the present work are in a
good agreement with the experimental data, which implies that the
present model ($\chi $QSM) produces reasonable form factors of
many different quantum numbers. We also predicted in that paper
the parity-violating asymmetries for the future G0 experiment at
backward angles. In the present paper we demonstrated that our
theoretical numbers reproduce also form factors from a combined
analysis of parity-violating electron scattering and $\nu$- and
$\bar{\nu}$- scattering. Altogether, comparing the results of the
$\chi$QSM with the overall world data one observes a remarkable
agreement. It seems that the Chiral Quark Soliton Model
($\chi$QSM), which has been applied over several years with an
accuracy of (10--30)\% to many observables of baryons in the octet
and decuplet, is formulated in terms of proper effective degrees
of freedom. The model is in fact the simplest quark model which
describes spontaneous breaking of chiral symmetry. It is based on
the $N\rightarrow \infty $ expansion of the QCD and appears to
describe the properties of the light baryons reasonably well.
Perhaps one can learn from the present comparison of the $\chi$QSM
with experiment that the degree of freedom of spontaneous breaking
of chiral symmetry governs not only the up- and down-sector of the
nucleon but also its strange quark content.

%
%==============================================================================
% Acknowledgement
%==============================================================================
%
\begin{acknowledgement}
The authors are grateful to Frank Maas for useful comments and discussions.
AS acknowledges partial financial support from Portugese Praxis
XXI/BD/15681/98. The work has also been supported by the Korean-German grant
of the Deutsche Forschungsgemeinschaft and KOSEF (F01-2004-000-00102-0). The
work is partially supported by the Transregio-Sonderforschungsbereich
Bonn-Bochum-Giessen, by the Verbundforschung of the Federal Ministry for
Education and Research, by the COSY-J{\"u}lich project, and by the EU
integrated infrastructure initiative ''Hadron Physics Project'' under
contract No. RII3-CT-2004-506078. The work of HCK is also supported by Korea
Research Foundation (Grant No. KRF-2003-070-C00015).
\end{acknowledgement}

%
%==============================================================================
% Bibliography
%==============================================================================
%

\end{document}